\documentclass[a4paper,11pt]{article}
\pdfoutput=1 % to allow compilation in JCAP
\usepackage{jcappub}
\usepackage{amsmath}
\usepackage{amssymb}
\usepackage{bm}
\usepackage{graphicx}
\usepackage{enumitem}
\usepackage{geometry}
\usepackage{xspace}
\usepackage{pdflscape}
\usepackage{enumerate}
\makeatletter
\gdef\@fpheader{}
\g@addto@macro\bfseries{\boldmath}
\makeatother

\newcommand{\ie}{{i.e.~}}

\newcommand{\eg}{e.g.~}

\newcommand{\etc}{\textsl{etc.~}}

\let\oldsqrt\sqrt
\def\sqrt{\mathpalette\DHLhksqrt}
\def\DHLhksqrt#1#2{%
\setbox0=\hbox{$#1\oldsqrt{#2\,}$}\dimen0=\ht0
\advance\dimen0-0.2\ht0
\setbox2=\hbox{\vrule height\ht0 depth -\dimen0}%
{\box0\lower0.4pt\box2}}

\newcommand{\order}[1]{\mathcal{O}\!\left(#1\right)}

\newcommand{\dd}{\mathrm{d}}
\newcommand{\ee}{e}

\newcommand{\sss}[1]{{\scriptscriptstyle{#1}}}
\newcommand{\boldmathsymbol}[1]{{\ensuremath{\boldsymbol{#1}}}}

\newcommand{\uPl}{\mathrm{Pl}}

\newcommand{\uinf}{\mathrm{inf}}

\newcommand{\uc}{\mathrm{c}}

\newcommand{\usssPl}{\sss{\uPl}}

\newcommand{\calH}{\mathcal{H}}
\newcommand{\calP}{\mathcal{P}}

\newcommand{\Rea}{\Re \mathrm{e}\,}
\newcommand{\Ima}{\Im \mathrm{m}\,}

\newcommand{\MeV}{\mathrm{MeV}}
\newcommand{\GeV}{\mathrm{GeV}}

\newcommand{\Mp}{M_\usssPl}

\newcommand{\efolds}{$e$-folds}

\newcommand{\beq}{\begin{equation}}
\newcommand{\eeq}{\end{equation}}
\newcommand{\bea}{\begin{equation}\begin{aligned}}
\newcommand{\eea}{\end{aligned}\end{equation}}

\newlength{\wsingfig}
\newlength{\wdblefig}
\newlength{\wquadfig}
\newlength{\wtriplefig}
\newcommand{\Eq}[1]{Eq.~(\ref{#1})}
\newcommand{\Eqs}[1]{Eqs.~(\ref{#1})}
\newcommand{\Fig}[1]{Fig.~{\ref{#1}}}

\newcommand{\Refa}[1]{Ref.~{\cite{#1}}}
\newcommand{\Refs}[1]{Refs.~{\cite{#1}}}
\newcommand{\Sec}[1]{Sec.~\ref{#1}}

\subheader{}

\title{On the choice of the collapse operator in cosmological Continuous
Spontaneous Localisation (CSL) theories}

\author[a]{J\'er\^ome Martin,}
\author[b,a]{Vincent Vennin}

\affiliation[a]{Institut d’Astrophysique de Paris, UMR 7095-CNRS,
Universit\'e Pierre et Marie Curie, 98\,bis boulevard Arago, 75014 Paris, France}
\affiliation[b]{Laboratoire Astroparticule et Cosmologie, CNRS, Universit\'e de Paris, 75013 Paris, France}

\emailAdd{jmartin@iap.fr}
\emailAdd{vincent.vennin@apc.in2p3.fr}

\date{today}

\begin{document}

\sloppy

\abstract{The Continuous Spontaneous Localisation (CSL) theory in the cosmological context is subject to uncertainties related to the choice of the collapse operator. In this paper, we constrain its form based on generic arguments. We show that, if the collapse operator is even in the field variables, it is unable to induce the collapse of the wavefunction. Instead, if it is odd, we find that only linear operators are such that the outcomes are distributed according to Gaussian statistics, as required by measurements of the cosmic microwave background. We discuss implications of these results for previously proposed collapse operators. We conclude that the cosmological CSL collapse operator should be linear in the field variables. 
}

%\keywords{inflation, physics of the early universe}

% \arxivnumber{XXXX.XXXXX}

\maketitle

\section{Introduction}
\label{sec:intro}

How a specific outcome is obtained when measuring a system placed in a quantum superposition~\cite{Bassi:2012bg} constitutes the measurement problem of quantum mechanics. In the cosmological context, its manifestation is particularly ``exacerbated''~\cite{Sudarsky:2009za}, since it is clear that no ``exterior'' observer can perform a measurement of the entire universe at early time that would make the wavefunction of cosmological structures collapse. This implies that the Copenhagen interpretation~\cite{1955mfqm.book.....V, Hartle:2019hae} cannot be used in this situation, and calls for alternative ``interpretations'', or rather formulations, of quantum mechanics.

One possibility is to consider that the standard quantum theory is only an approximation of a more general framework. Dynamical collapse models~\cite{Ghirardi:1985mt, Diosi:1988uy, Ghirardi:1989cn, Bassi:2003gd, Bassi:2012bg} follow this reasoning and introduce
a non-linear and stochastic modification to the Schr\"odinger
equation. The most refined version of these models is the Continuous
Spontaneous Localisation (CSL) theory~\cite{Ghirardi:1989cn}, in which the  modified Schr\"odinger equation reads
\bea 
\label{eq:csl:real}
\frac{\dd\left\vert\Psi\right\rangle}{\dd\eta} = & \left\lbrace -i \int \dd \bm{x} \hat{\calH}(\bm{x})+\frac{\sqrt{\gamma}}{m_0}\int \dd \bm{x} \left[\hat{C}(\bm{x}) - \left\langle \hat{C}(\bm{x}) \right\rangle \right]\xi(\eta,\bm{x})
\right. \\ & \left. 
-\frac{\gamma}{2 m_0^2} \int \dd\bm{x} \, \dd \bm{y} 
\left[\hat{C}(\bm{x}) - \left\langle \hat{C}(\bm{x}) \right\rangle \right]
G(\bm{x}-\bm{y})
\left[\hat{C}(\bm{y}) - \left\langle \hat{C}(\bm{y}) \right\rangle \right]
\right\rbrace \left\vert\Psi\right\rangle,
\eea 
where $\xi(\eta,\bm{x})$ is a white Gaussian noise with two-point correlation function given by
\bea
\mathbb{E}\left[\xi(\eta,\bm{x}) \, \xi(\eta',\bm{y})\right] = G(\bm{x}-\bm{y})\delta(\eta-\eta')\, .
\eea 
In these expressions, $G(\bm{x}-\bm{y})$ is a Gaussian smearing function over the distance $r_\uc$, which is the first free parameter of the theory,
\bea 
G(\bm{x}-\bm{y})=\frac{\ee^{-\frac{\left\vert\bm{x}-\bm{y} \right\vert^2}{4r_\uc^2}}}{\left(4\pi r_\uc^2\right)^{3/2}}\, .
\label{eq:G:Gaussian:def}
\eea 
In \Eq{eq:csl:real}, $\vert \Psi \rangle$ is the wavefunction of the system under consideration. In standard quantum mechanics, it evolves with the local Hamiltonian $\hat{H}=\int \dd \bm{x} \hat{\calH}(\bm{x})$, which gives rise to the first term in the right-hand side of \Eq{eq:csl:real}. The two additional terms are controlled by $\gamma$, which is the second free parameter of the theory, and $m_0$, which is a reference mass (usually the mass of a nucleon). They involve a collapse operator $\hat{C}(\bm{x})$ that we keep generic for the moment, and \Eq{eq:csl:real} induces the collapse of the wavefunction towards one of the eigenstates of this operator. Which eigenstate is selected depends on the realisation of the stochastic process $\xi(\eta,\bm{x})$. 

Owing to the presence of $\langle \hat{C}(\bm{x}) \rangle=\langle \Psi \vert  \hat{C}(\bm{x})\vert \Psi \rangle$, the modified Schr\"odinger equation is non linear in $\vert \Psi \rangle$, which enables the breakdown of quantum superpositions. The stochasticity is then necessary to prevent faster-than-light signalling~\cite{Bassi:2012bg}, so the structure of the modification is essentially unique. It is also worth stressing out that it is naturally endowed with an amplification mechanism, which
allows microscopic systems to be described by the standard rules of
quantum mechanics to a good accuracy, while preventing macroscopic systems from being found in a superposition of macroscopically distinct configurations.

Since macroscopic objects are always found to be localised in space, it can be argued~\cite{Bassi:2003gd} that a natural choice for the collapse operator is the mass density operator
\bea
\label{eq:C:mass:density}
\hat{C}(\bm{x}) = \sum_i m_i \hat{a}^\dagger_i (\bm{x})\hat{a}_i (\bm{x}),
\eea 
where $\hat{a}^\dagger_i (\bm{x})$ and $\hat{a}_i (\bm{x})$ are the creation and annihilation operators of a particle of type $i$, which has mass $m_i$, at location $\bm{x}$. One may be concerned with the fact that in practice, one can measure other microscopic properties than the spatial position of particles (say their spin). The argument is however that even in such situations, ultimately, one only observes the spatial position of macroscopic objects (say the direction of an arrow in a detector, the position of a dot on a screen, \etc). 

With such a choice for the collapse operator, the two
parameters $\gamma$ and $r_\uc$ have been constrained in various
laboratory experiments. The strongest bounds so far come from $X$-ray
spontaneous emission~\cite{2015arXiv150205961C}, force noise
measurements on ultracold cantilevers~\cite{2016PhRvL.116i0402V}, and
gravitational-wave interferometers~\cite{Carlesso:2016khv}. These
constraints leave the region of parameter space around
$r_\uc \sim 10^{-8}-10^{-4}\mathrm{m}$ and
$\lambda\sim 10^{-18}-10^{-10}\mathrm{s}^{-1}$ viable, where
$\lambda\equiv \gamma/(8\pi^{3/2}r_\uc^3)$ is known as the collapse rate.

Since the typical physical scales involved in cosmology are many orders of magnitude different from those encountered in the lab (in the early universe, the relevant energy scales can be as high as $\sim 10^{15}\GeV$, corresponding to densities of $\sim 10^{80}\, \mathrm{g}\times \mathrm{cm}^{-3}$), cosmological observations, which have reached an exquisite level of precision \eg in measurements of the Cosmic Microwave Background (CMB) anisotropies~\cite{Aghanim:2018eyx}, may lead to competitive constraints. Moreover, as argued above, it is essential to understand how cosmological structures, born from quantum fluctuations in the early universe, collapse into a specific configuration before the CMB is emitted. For these two reasons, it is interesting to try and apply the CSL theory to primordial cosmological fluctuations.

When doing so, one however faces the issue that while CSL is designed as a non-relativistic theory for quantum particles, cosmological fluctuations are to be described by a quantum scalar field $\hat{v}(\bm{x})$~\cite{Mukhanov:1981xt,Kodama:1985bj}, evolving on an expanding (hence curved) geometrical background. Nonetheless, since this background is statistically homogeneous and isotropic, each Fourier mode of this field evolves independently in the standard theory, and follows the equation of a single-particle quantum parametric oscillator. In spite of this simplification, the issue of which collapse operator needs to be used in this context still remains and several choices are a priori possible.

In \Refa{Martin:2019jye}, see also \Refs{Martin:2019oqq, Martin:2020sdm}, it was proposed that a natural extension of the notion of mass density to fields is given by the energy density $\rho$,
\bea 
\label{eq:C:energy:density}
\hat{C}(\bm{x}) = \hat{\rho}(\bm{x}) .
\eea 
Since $\hat{v}(\bm{x})$ only self interacts gravitationally, ``energy'' is indeed to be understood as ``mass energy'' here.
In General Relativity, the energy density however depends on the choice of hypersurface on which it is measured. It was found that except for the special case where the energy density is evaluated in the comoving threading, or unless inflation proceeds at very low energy, for the values of $\lambda$ and $r_\uc$ mentioned above, the CSL corrections are incompatible with current measurements of the CMB. This means that when embedding CSL in a more fundamental, relativistic theory, a mechanism that selects out the comoving threading must emerge, or a strong running of the effective value of $\lambda$ and $r_\uc$ with the energy scale should be obtained; or some other specific mechanism must be found. In any case, this shows the potential of cosmology to guide and constrain possible relativistic generalisations of the CSL theory, which is the main conclusion of \Refa{Martin:2019jye}.

A fundamental difference between \Eqs{eq:C:mass:density} and~\eqref{eq:C:energy:density} is that, while \Eq{eq:C:mass:density} is quadratic in the creation and annihilation operators, the proposal~\eqref{eq:C:energy:density} also contains linear contributions. The reason is that the energy density can be expanded as
\bea 
\hat{\rho}(\bm{x}) = \bar{\rho} + \widehat{\delta\rho}\left(\bm{x}\right),
\eea 
where $ \bar{\rho}$ is the homogeneous, classical component of the energy density, which gives a vanishing contribution in the additional terms of \Eq{eq:csl:real}. The fluctuation in the energy density, $\widehat{\delta\rho}(\bm{x})$, is a (non-linear) function of $\hat{v}(\bm{x})$ and its conjugated momentum $\hat{p}(\bm{x})$, hence of the ladder operators. In cosmological perturbation theory, the dominant contribution however comes from the linear terms, which differs from \Eq{eq:C:mass:density} where such linear contributions are absent.

This property was viewed by \Refa{Gundhi:2021dkh} as something which makes the choice of the energy density unrealistic and this motivated \Refa{Gundhi:2021dkh} to propose another collapse operator when using CSL in the cosmological context, namely the free Hamiltonian density of the system,
\bea 
\label{eq:C:H}
\hat{C}(\bm{x}) = \hat{\calH}(\bm{x}).
\eea 
It was then argued that, with this choice, the corrections to the power spectrum of the fluctuations are tiny, and that no competitive constraints can be obtained from cosmology. In other words, the strong cosmological constraints on CSL found in \Refs{Martin:2019oqq, Martin:2020sdm} would just be artefacts of an unnatural choice for the collapse operator.

The goal of this paper is to further discuss the choice of the CSL collapse operator in a cosmological context. In the course of this general study, we will further investigate the possibility to consider the Hamiltonian density as the collapse operator, and we will re-examine whether it succeeds in making the wavefunction of cosmological perturbations collapse towards the state observed in the CMB. The article is organised as follows. In \Sec{sec:cslinflation}, we show how the CSL Schr\"odinger equation~\eqref{eq:csl:real} can be employed to describe inflationary fluctuations. In \Sec{subsec:modifiedS}, we write this equation in Fourier space. In \Sec{subsec:collapseoperator}, we introduce in more detail the proposal made in \Refa{Gundhi:2021dkh}, and compare it with the one of \Refs{Martin:2019jye, Martin:2019oqq, Martin:2020sdm}. In \Sec{sec:power}, we turn to the calculation of the power spectrum of cosmological fluctuations. In \Sec{subsec:define}, we discuss how the cosmological power spectrum can be defined in CSL. In \Sec{subsec:collapse}, we introduce the quantity denoted $R(k)$, which allows us to track whether or not the wavefunction has actually collapsed. In \Sec{subsec:calculationps}, we discuss which collapse operators are able, at the same time, to make the wavefunction collapse and lead to a power spectrum that is in agreement with cosmological data. In particular, we show that the collapse operator of \Refa{Gundhi:2021dkh} is unable to make the wavefunction of cosmological structures collapse, 
so it cannot explain the emergence of cosmological structures in the early universe. This is due to the invariance of the collapse operator under sign flipping of the field variables, and implies that the power spectrum $\calP_v(k)$ strictly vanishes in this model. In \Sec{sec:NG}, we show how higher correlation functions, namely non-Gaussianity, can help us to further constrain the form of the collapse operator. In this section, a full solution of the modified Schr\"odinger equation is presented in the case where the collapse operator is a linear function of the field variables, and where all parameters of the wavefunction can be expressed in terms of the solutions to a single ordinary, linear differential equation. Finally, in \Sec{sec:Discussion}, we present our conclusions. 
\section{CSL and inflation}
\label{sec:cslinflation} 
In this section, we first explain how the CSL Schr\"odinger equation~\eqref{eq:csl:real}, written in real space, can be written in Fourier space. We compare the approaches of \Refs{Martin:2019jye} and \cite{Gundhi:2021dkh} and check that they use the same stochastic Schr\"odinger equation, thus confirming that the only difference between those two articles lies in the choice of the collapse operator, see \Eqs{eq:C:energy:density} and~\eqref{eq:C:H} respectively. In the context of cosmology, we then show that an important difference between the two proposals is that, while \Eq{eq:C:energy:density} leaves different Fourier modes uncoupled, \Eq{eq:C:H} leads to explicit mode coupling.

\subsection{The modified Schr\"odinger equation}
\label{subsec:modifiedS}

The modified Schr\"odinger equation~\eqref{eq:csl:real} applies to the wavefunctional of the real scalar field $v(\bm{x})$, the so-called Mukhanov-Sasaki variable, which we denote $\Psi[v(\bm{x})]$. This can also be written as a wavefunctional of all Fourier modes of the field $v(\bm{x})$, \ie \begin{align} 
\label{eq:Psi:Real:Fourier}
\Psi\left[\left\lbrace v(\bm{x}) \right\rbrace; \bm{x}\in \mathbb{R}^{3}\right]=\Psi\left[\left\lbrace v^s_{\bm{k}}(\eta)\right\rbrace ;\bm{k}\in \mathbb{R}^{3+},s\in \lbrace \mathrm{R,I} \rbrace\right], 
\end{align}
where $s$ labels the real and imaginary parts of $v_{\bm k}(\eta)$. Let us notice that \Refa{Gundhi:2021dkh} considers the curvature perturbation ${\cal R}_{\bm k}$ rather than the Mukhanov-Sasaki variable. Both approaches are in fact similar 
since these two quantities only differ by a background quantity, namely ${\cal R}_{\bm k}=v_{\bm k}/z$, with $z=a\Mp\sqrt{2\epsilon_1}$, $a$ being the cosmological scale factor, $\epsilon_1$ the first slow-roll parameter and $\Mp$ the reduced Planck mass. In Fourier space, the CSL equation~\eqref{eq:csl:real} reads
\bea 
\label{eq:csl:Fourier}
\frac{\dd\left\vert\Psi\right\rangle}{\dd\eta} = & \int_{\mathbb{R}^3} \dd\bm{k} \left\lbrace -i \hat{\calH}({\bm{k}})+\frac{\sqrt{\gamma}}{m_0} \left[\hat{C}^\dagger_{\bm{k}} - \left\langle \hat{C}^\dagger_{\bm{k}} \right\rangle \right]\xi_{\bm{k}}(\eta)
\right. \\ & \left. 
-\frac{\gamma}{2 m_0^2} 
\left[\hat{C}^\dagger_{\bm{k}} - \left\langle \hat{C}^\dagger_{\bm{k}} \right\rangle \right]
(2\pi)^{3/2}G_{\bm{k}}
\left[\hat{C}_{\bm{k}} - \left\langle \hat{C}_{\bm{k}} \right\rangle \right]
\right\rbrace \left\vert\Psi\right\rangle.
\eea 
In this equation, $\hat{C}_{{\bm k}}$ denotes the Fourier transform of the collapse operator, $\hat{C}({\bm x})=(2\pi)^{-3/2}\int \dd {\bm k}\, \hat{C}_{{\bm k}}\, e^{-i{\bm k}\cdot {\bm x}}$, while $\xi_{{\bm k}}(\eta)$ and $G_{{\bm k}}$ are the Fourier transforms of the noise and the smearing function, respectively. The two-point correlation function of the noise can be expressed as
\bea 
\mathbb{E}\left[\xi_{\bm{k}}(\eta)\, \xi^*_{\bm{k}'}(\eta')\right]
=4\pi
\delta(\eta-\eta') \delta(\bm{k}-\bm{k}')
\int _0^{+\infty} \dd r \, r^2 G(r)\, \mathrm{sinc}(kr)
\, .
\eea
If one chooses a Gaussian smearing function, as done in \Eq{eq:G:Gaussian:def}, then $G_{\bm{k}}=(2\pi)^{-3/2}e^{-k^2r_\uc^2}$ [and the factor $(2\pi)^{3/2}$ appearing in front of $G_{{\bm k}}$ in \Eq{eq:csl:Fourier} cancels out] and the noise correlation function becomes 
\bea 
\mathbb{E}\left[\xi_{\bm{k}}(\eta)\, \xi^*_{\bm{k}'}(\eta')\right]
=    
e^{-k^2\, r_\uc^2}
\delta(\eta-\eta') \delta(\bm{k}-\bm{k}')
\, .
\eea
These are the equations used in \Refa{Gundhi:2021dkh}.
It is also possible to implement the smearing directly in the 
collapse operator rather than introducing a function $G({\bm x}-{\bm y})$ (or $G_{\bm k}$) in the CSL equation. This was the route chosen in Ref.~\cite{Martin:2019jye}. Indeed, one can define a new collapse operator $\hat{\bar{C}}_{{\bm k}}\equiv e^{-k^2r_\uc^2/2}\hat{C}_{{\bm k}}$
and a new noise, $\bar{\xi}_{{\bm k}}(\eta)\equiv   e^{k^2r_\uc^2/2}\xi_{{\bm k}}(\eta)$ and, with these new definitions, the CSL equation~\eqref{eq:csl:real} takes the form
\bea 
\label{eq:csl:Fourier2}
\frac{\dd\left\vert\Psi\right\rangle}{\dd\eta} = & \int_{\mathbb{R}^3} \dd\bm{k} \left\lbrace -i \hat{\calH}({\bm{k}})+\frac{\sqrt{\gamma}}{m_0} \left[\hat{\bar{C}}^\dagger_{\bm{k}} - \left\langle \hat{\bar{C}}^\dagger_{\bm{k}} \right\rangle \right]\bar{\xi}_{\bm{k}}(\eta)
\right. \\ & \left. 
-\frac{\gamma}{2 m_0^2} 
\left[\hat{\bar{C}}^\dagger_{\bm{k}} - \left\langle \hat{\bar{C}}^\dagger_{\bm{k}} \right\rangle \right]
\left[\hat{\bar{C}}_{\bm{k}} - \left\langle \hat{\bar{C}}_{\bm{k}} \right\rangle \right]
\right\rbrace \left\vert\Psi\right\rangle,
\eea 
with
\bea 
\mathbb{E}\left[\bar{\xi}_{\bm{k}}(\eta)\, \bar{\xi}^*_{\bm{k}'}(\eta')\right]
=    
\delta(\eta-\eta') \delta(\bm{k}-\bm{k}')
\, .
\eea
This is the CSL equation used in \Refa{Martin:2019jye} and we conclude that it is strictly equivalent with the one employed in \Refa{Gundhi:2021dkh}. Notice that the smearing procedure adopted in \Refa{Martin:2019jye}, which represents an essential part of what the CSL equation is about, was criticised in~\Refs{Bengochea:2020efe} and~\cite{Bengochea:2020qsd} (those remarks were answered in \Refs{Martin:2019oqq} and~\cite{Martin:2020sdm}). The above result shows that these criticisms would also apply to \Refa{Gundhi:2021dkh} since the smearing procedures in \Refs{Martin:2019jye} and~\cite{Gundhi:2021dkh} are in fact identical. The only difference between those two articles is therefore the choice of the collapse operator, a question on which we focus in the rest of this paper.

Let us also discuss the quantity $\hat{\calH}({\bm{k}})$ appearing in \Eqs{eq:csl:Fourier} and \eqref{eq:csl:Fourier2}. This quantity is the Hamiltonian when expressed in terms of the Fourier transform of the field variables. At leading order in cosmological perturbation theory, the Hamiltonian is quadratic in the field variables, and in general it can be written in matricial form as
\bea 
\label{eq:Hamiltonian:quadratic}
\hat{H} = \int \dd \bm{x}\, \hat{\mathcal{H}}(\bm{x}) = 
\frac{1}{2} \int \dd \bm{x} \,
 \hat{\boldmathsymbol{z}}^\mathrm{T}(\bm{x}) \boldmathsymbol{H}(\eta) \hat{\boldmathsymbol{z}}(\bm{x}),
\eea 
where $\boldmathsymbol{z}$ is a vector containing the Mukhanov-Sasaki field and its conjugated momentum, and $\boldmathsymbol{H}$ is a time-dependent two-by-two matrix, which involves functions of the background geometry and gradient operators. This gives rise to
\bea 
\hat{H} = 
\int_{\mathbb{R}^3}\dd\bm{k}\, \hat{\calH}({\bm{k}})
= \int_{\mathbb{R}^{3+}}\dd \bm{k}\sum_{s=\mathrm{R,I}} \hat{\mathcal{H}}^s({\bm{k}})\, ,
\eea 
with 
\bea
\label{eq:HamiltonianFourier}
\hat{\mathcal{H}}^\mathrm{R}({\bm{k}}) = \Rea\left[\hat{\boldmathsymbol{z}}^\mathrm{T}(\bm{k})\right]\boldmathsymbol{H}(\eta)\Rea\left[\hat{\boldmathsymbol{z}}(\bm{k})\right]
\eea 
and similarly for $\hat{\mathcal{H}}^\mathrm{I}({\bm k})$, where the gradients in $\boldmathsymbol{H}$ are replaced by the relevant functions of $\bm{k}$ [so, strictly speaking, one should introduce different notations for $\boldmathsymbol{H}(\eta)$ in \Eqs{eq:Hamiltonian:quadratic} and~(\ref{eq:HamiltonianFourier})]. The operator $\hat{H}$ is thus separable in Fourier space, since it can be written as a sum of operators acting in each Fourier subspace separately. It gives rise to the standard Hamiltonian evolution of the system, which does not mix different wavenumbers. 

\subsection{The collapse operator}
\label{subsec:collapseoperator}

The other quantity of interest appearing in \Eq{eq:csl:Fourier} is $\hat{C}_{\bm{k}}$ (or $\hat{\bar{C}}_{\bm k}$), which, as already mentioned, is the Fourier transform of $\hat{C}(\bm{x})$, the collapse operator. In the context of cosmology and cosmic inflation, the main discussion has been about the choice of this operator and different possibilities have recently been discussed in the litterature~\cite{Martin:2019jye,Bengochea:2020efe,Bengochea:2020qsd,Gundhi:2021dkh}. In Ref.~\cite{Martin:2019jye}, it was proposed that a natural choice is the energy density, see~\Eq{eq:C:energy:density}. Then, as argued above, at leading order in cosmological perturbation theory, this implies that the collapse operator is linear in field variables and can be written as
\bea
\label{eq:C:linear}
\hat{\bar{C}}_{\bm{k}} = \alpha_{\bm k}(\eta) \hat{v}_{\bm{k}} + \beta_{\bm k}(\eta) \hat{p}_{\bm{k}}\, .
\eea 
Let us notice that the same formula could have been written for $\hat{{C}}({\bm k})$ depending of whether the smearing exponential $e^{-k^2r_\uc^2}$ is included or not in the coefficients $\alpha_{\bm k}(\eta)$ and $\beta_{\bm k}(\eta)$. In any case, this means that the additional terms in the Schr\"odinger equation are separable in Fourier space since they can be written as a sum of operators acting in each Fourier subspace separately. Therefore, even in the presence of the additional CSL terms, the wavefunction remains factorisable,
\bea 
\label{eq:Psi:Fourier}
\Psi\left[\left\lbrace v(\bm{x}) \right\rbrace; \bm{x}\in \mathbb{R}^{3}\right]=\prod _{{\bm k}\in\mathbb{R}^{3+};s=\mathrm{R,I}}\Psi_{\bm k}^s(v_{\bm k}^s),
\eea 
if its initial state is so (which is the case for the Bunch-Davies vacuum~\cite{Bunch:1978yq} used in cosmology).
This is why, in the approach of \Refa{Martin:2019jye}, mode coupling can only appear at next-to-leading order in perturbation theory (\ie from the contribution to the energy density that is quadratic in field variables, as well as the -- standard -- contribution to the free Hamiltonian that is cubic in field variables). As a result, \Eq{eq:csl:Fourier2} can be re-expressed as a collection 
of equations for each Fourier mode, namely
\bea 
\label{eq:csl:Fourierseparable}
\frac{\dd\left\vert\Psi_{\bm k}^s\right\rangle}{\dd\eta} = & \left\lbrace -i \hat{\calH}^s({\bm{k}})+\frac{\sqrt{\gamma}}{m_0} \left[\hat{\bar{C}}^s_{\bm{k}} - \left\langle \hat{\bar{C}}^s_{\bm{k}} \right\rangle \right]\bar{\xi}_{\bm{k}}^s(\eta)
%\right. \\ & \left. 
-\frac{\gamma}{2 m_0^2} 
\left[\hat{\bar{C}}^s_{\bm{k}} - \left\langle \hat{\bar{C}}^s_{\bm{k}} \right\rangle \right]^2
\right\rbrace \left\vert\Psi_{\bm k}^s\right\rangle,
\eea 
which is exactly Eq.~(18) of \Refa{Martin:2019jye}.\footnote{Since the wavenumbers $k$ in \Eqs{eq:csl:Fourier}, (\ref{eq:csl:Fourier2}) and~(\ref{eq:csl:Fourierseparable}) are physical wavenumbers, in an expanding universe, additional powers of the scale factor $a(\eta)$ appear in the different terms of those equations when expressed in terms of comoving wavenumbers, as done in \Refa{Martin:2019jye}.}

As mentioned above, after \Refa{Martin:2019jye}, discussions about the choice of the collapse operators were published. In particular, in \Refs{Bengochea:2020efe} and~\cite{Bengochea:2020qsd}, it was 
argued that more complicated collapse operators, such as $T_{\mu}^{\mu}$, $(T_{\mu \nu}T^{\mu \nu})^{1/2}, \cdots $ (where $T_{\mu\nu}$ is the stress-energy tensor), are also possible and could potentially modify the conclusions of \Refa{Martin:2019jye}. However, it was shown in \Refa{Martin:2020sdm} that all these alternatives reduce to \Eq{eq:C:linear} with slight modifications of order one in the coefficients $\alpha_{\bm k}$ and $\beta_{\bm k}$, hence they cannot substantially modify the conclusions based on the choice~(\ref{eq:C:linear}). This may not come as a surprise since the matrix element of these operators are all of the order of the energy density during inflation. 

More recently, \Refa{Gundhi:2021dkh} followed a similar line of reasoning and also proposed yet another collapse operator corresponding to the choice~\eqref{eq:C:H}. This choice implies that $\hat{C}_{\bm{k}} = \hat{\calH}_{\bm{k}}$, a quantity which should not be confused with $\hat{\calH}({\bm{k}})$, and which, under the quadratic assumption~\eqref{eq:Hamiltonian:quadratic}, reads
\bea
\label{eq:H_k}
\hat{\mathcal{H}}_{\bm{k}} &= \frac{1}{2} (2\pi)^{-3/2} \int_{\mathbb{R}^{3}} \dd \bm{q} \hat{\boldmathsymbol{z}}^\mathrm{T} (\bm{q}) \boldmathsymbol{H}(\eta) \hat{\boldmathsymbol{z}}(\bm{k}-\bm{q})\, .
\eea 
This operator is manifestly not separable in Fourier space, and leads to mode coupling. It implies that the evolution of different Fourier modes is not independent in the CSL dynamics, and that the quantum state can no longer be factorised according to \Eq{eq:Psi:Fourier}. In \Refa{Gundhi:2021dkh}, it is argued that this 
property 
is what makes this choice more natural and better justified as it would be a way ``{\it to retain the characteristic traits}'' of CSL when generalised to a cosmological situation.
 In other words, because the collapse operator is usually taken to be quadratic in the creation and annihilation operators, see \Eq{eq:C:mass:density}, this should also be the case when dealing with cosmic inflation, and this is not what is done in \Refa{Martin:2019jye}. 
We will further examine this claim below, but in order to avoid possible confusion, let us stress already that the form of the collapse operator has of course nothing to do with the fact that CSL is a non-linear theory, which is required to let it break the superposition principle. Indeed, even if the collapse operator is linear in the field variables, the additional CSL terms are always non linear in the wavefunction, which is what matters.

One can of course discourse about the advantages and disadvantages of linear and non-linear collapse operators and which ones are more natural. The crucial test is however to determine whether or not they can lead to theories that are phenomenologically acceptable, \ie whether or not they can meet two requirements: (i) they must make the wavefunction collapse, so as to explain the emergence of the classical structures we observe; and (ii) the statistical distribution of the collapsed states must be in agreement with observations. This is the case if it follows the Born rule since the ``standard'' calculation is known to provide an excellent fit to the CMB data.

In \Refa{Gundhi:2021dkh}, it is claimed that the proposal~\eqref{eq:C:H} meets the second requirement (while the first one is not explicitly examined), since the CSL terms only provide tiny corrections to the predicted power spectrum, making the CMB unable to provide competitive constraints on CSL in that case. In the following sections, we will argue that, in fact, neither requirement is met by the choice~\eqref{eq:C:H}, making the theory unsuitable for cosmology.

\section{Power Spectrum}
\label{sec:power}

\subsection{How to define the power spectrum in CSL?}
\label{subsec:define}
\begin{figure} 
    \centering
    \includegraphics[width=1\textwidth]{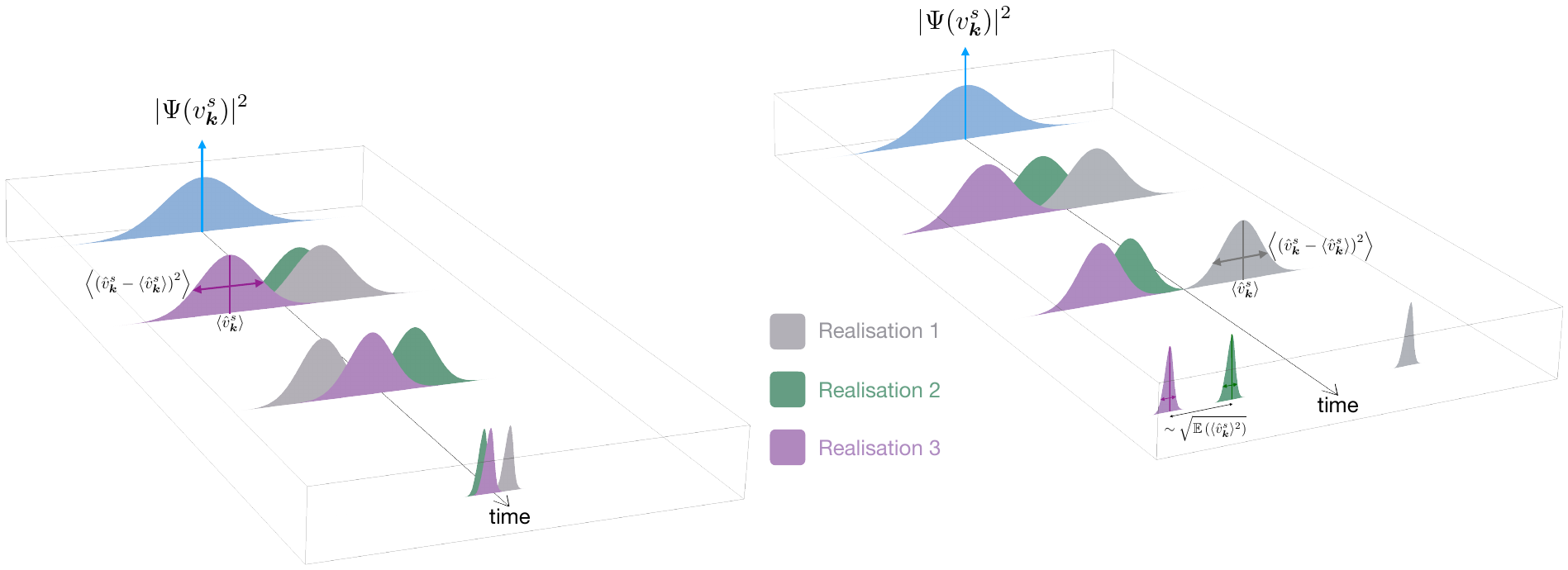}
    \caption{In the framework of CSL, the wave-function of cosmological perturbations, $\Psi(v_{\bm k}^s)$, is a stochastic quantity. As a consequence, its quantum mean value $\langle \hat{v}_{\bm k}^s\rangle$ and quantum dispersion $\langle (\hat{v}_{\bm k}^s-\langle \hat{v}_{\bm k}^s\rangle)^2\rangle $ are random variables. In the left panel, we have sketched the stochastic ``trajectories'' of this wave-function for three different realisations. At the final time, the
  dispersion of the means, $\mathbb{E}[\langle \hat{v}_{\bm
      k}^s\rangle^2]$ (the stochastic average of the means vanishes
  $\mathbb{E}[\langle \hat{v}_{\bm k}^s\rangle]=0$) is not small
  compared to the width of the wave-functions, $\mathbb{E}[\langle
    (\hat{v}_{\bm k}^s-\langle \hat{v}_{\bm
      k}^s\rangle)^2\rangle]$, and our collapse criterion is not
  satisfied. In this case, the different wave-functions representing
  different realisations are not sufficiently separated to account for
  the emergence of different outcomes. In the right panel, on the
  contrary, our criterion is satisfied and different realisations do
  correspond to well-separated outcomes. Figure reproduced from \Refa{Martin:2020sdm}.
 \label{fig:realisations}}
\end{figure}
Let us consider a given wavenumber $\bm{k}$, for which $v_{\bm{k}}^s$ is being measured on the CMB temperature and polarisation anisotropies maps. Along a given realisation of the stochastic equation~\eqref{eq:csl:Fourier}, the quantum expectation value of $v_{\bm{k}}^s$, $\langle \hat{v}^s_{\bm{k}} \rangle$, and the quantum expectation value of its variance, $\langle (\hat{v}^s_{\bm{k}} - \langle \hat{v}^s_{\bm{k}} \rangle)^2\rangle$, evolve in a stochastic way.\footnote{If the collapse operator is linear in the field variables, as in \Eq{eq:C:linear}, the variance turns out to follow a deterministic equation of motion, see \Eq{eq:eom:Riccatti:Omegak} below, and only the first moment is stochastic.\label{footnote:Deterministic:Variance}} In the limit where the state is fully collapsed, the variance of the wavefunction vanishes, and the (squared) wavefunction becomes a Dirac distribution centered on $\langle \hat{v}^s_{\bm{k}} \rangle$. It is important to emphasise that, in the standard (\ie non-CSL) picture, the quantity $\langle \hat{v}^s_{\bm{k}} \rangle$ remains zero. On the contrary, in CSL, due to the stochastic evolution of the wavefunction, we expect each realisation to acquire a non-vanishing $\langle \hat{v}^s_{\bm{k}} \rangle$. Once the wavefunction has collapsed around $\langle \hat{v}^s_{\bm{k}} \rangle$, the quantity that is measured can be nothing else than $\langle \hat{v}^s_{\bm{k}} \rangle$, so observations give access to the stochastic distribution associated to the first moment of the wavefunction. In particular, the power spectrum, \ie the two-point correlation function observed in the CMB map, is defined as the second moment of that distribution, namely
\bea
\label{eq:def:PowerSpectrum}
P_v(k) \propto \mathbb{E}\left(\left\langle \hat{v}_{\bm{k}}^s\right\rangle ^2\right)-\mathbb{E}^2\left(\left\langle
\hat{v}_{\bm{k}}^s\right \rangle\right)\, .
\eea 
As will be shown below explicitly, because of the invariance of the averaged theory under sign flipping of the field variables, one has $\mathbb{E}\left(\left\langle
\hat{v}_{\bm{k}}^s\right \rangle\right)=0$ and, as a consequence, the power spectrum reduces to 
\begin{align}
\label{eq:def:PowerSpectrum:2}
P_v(k) \propto \mathbb{E}\left(\left\langle \hat{v}_{\bm{k}}^s\right\rangle ^2\right).
\end{align}
This stochastic expectation value then needs to be compared with the two-point function of cosmological fluctuations measured on the CMB map~\cite{Leon:2011ca,Leon:2015eza}, where the identification between stochastic and spatial averages can be made in the ergodic limit, the deviation from which quantifies cosmic variance~\cite{Grishchuk:1997pk}.

Let us note that, in quantum mechanics, observables can be attached to collapsed systems only. In the Copenhagen interpretation, this is because the measurement process itself does not leave the state uncollapsed, while in CSL, this is because the collapse is dynamically realised before the measurement is complete. As a consequence, the power spectrum is unambiguously defined for collapsed states only. With the definition we have adopted, it is interesting to notice that, if $\gamma \rightarrow 0$ (or, equivalently, $\lambda \rightarrow 0$), that is to say in the absence of the CSL extra terms, the power spectrum necessarily vanishes, namely 
\begin{align}
    \lim_{\gamma \rightarrow 0}P_v(k)=0,
\end{align}
because, as mentioned before, $\langle \hat{v}^s_{\bm{k}} \rangle$ remains zero in that case. This is in agreement with the idea that no structure is formed (hence the power spectrum vanishes) in the absence of collapse, and this is the reason why this way of defining the power spectrum seems intuitive.
However, for practical purposes, any other ``definition'' of the power spectrum that matches \Eq{eq:def:PowerSpectrum} in the collapsed limit, but that may differ when evaluated on uncollapsed states, can a priori also be used.
This is for instance the case for the quantity $\mathbb{E}\left(\left\langle \hat{v}_{\bm{k}}^s{}^2\right\rangle \right)$, that is interpreted as the power spectrum probed 
by CMB experiments in \Refa{Gundhi:2021dkh}. In the fully collapsed state, when measuring $\hat{v}_{\bm{k}}^s{}^2$, one necessarily obtains $\left\langle \hat{v}_{\bm{k}}^s\right\rangle ^2$, hence this coincides with \Eq{eq:def:PowerSpectrum:2}. 

This quantity is perturbatively calculated in \Refa{Gundhi:2021dkh}, see the considerations around Eqs.~(17)-(18) and Eqs.~(94)-(95), and it is found that, if the collapse operator is taken to be the free Hamiltonian~(\ref{eq:C:H}), then the corresponding CSL corrections are negligible. This leads \Refa{Gundhi:2021dkh} to conclude that the predictions of CSL are close to that of the standard Copenhagen version of the theory. Let us stress that for this to be correct, one first needs to check that the wavefunction has indeed collapsed, otherwise the power spectrum is simply not defined. The validity of the conclusion made in \Refa{Gundhi:2021dkh} is therefore subject to the efficacy of the collapse, which is not discussed in that reference, but that we now study.

\subsection{Collapse of the wavefunction}
\label{subsec:collapse}

We have seen before that, in order to properly describe inflationary perturbations, CSL must be such that the power spectrum (and higher correlation functions, see \Sec{sec:NG} below) is in agreement with observations, and such that the wavefunction collapses. In this section, we study this second requirement.

In practice, collapse is never fully achieved (one never reaches a Dirac wavefunction exactly), and must be decided upon a certain criterion that can be introduced as follows. Let us consider two realisations of the CSL equation~\eqref{eq:csl:Fourier}, that we label ``1'' and ``2''. For the two states they evolve into to be properly resolved, the distance between their means should be larger than the sum of their standard deviation, 
\bea 
\label{eq:collapse:crit:1}
\left \vert  \left\langle \hat{v}^s_{\bm{k}} \right\rangle_1- \left\langle \hat{v}^s_{\bm{k}} \right\rangle_2 \right\vert \gg \sqrt{\left\langle (\hat{v}^s_{\bm{k}} - \langle \hat{v}^s_{\bm{k}} \rangle)^2 \right\rangle_1}+\sqrt{\left\langle (\hat{v}^s_{\bm{k}} - \langle \hat{v}^s_{\bm{k}} \rangle)^2 \right\rangle_2}\, . 
\eea 
The situation is sketched in \Fig{fig:realisations} (for display convenience, the case without mode coupling is depicted, where each component $\Psi_{\bm{k}}^s$ of \Eq{eq:Psi:Fourier} can be treated separately, but the argument is generic). Whether \Eq{eq:collapse:crit:1} is satisfied or not depends on the pair of trajectories one is considering.
One may however require that it is satisfied for \emph{most} pairs of realisations. The \emph{typical} squared distance between the first moments is given by the stochastic average of the distance away from the mean, $\mathbb{E}\lbrace \left[\langle \hat{v}^s_{\bm{k}} \rangle - \mathbb{E} (\langle \hat{v}^s_{\bm{k}} \rangle) \right]^2 \rbrace$, while the typical second moment is given by $\mathbb{E}[\langle(\hat{v}_{\bm k}^s-\langle \hat{v}_{\bm k}^s\rangle)^2\rangle]$. At initial time, one has $\langle \hat{v}^s_{\bm{k}} \rangle=0$, so $\mathbb{E} (\langle \hat{v}^s_{\bm{k}} \rangle)=0$ at any time if the averaged theory is invariant by flipping the sign of $v(\bm{x})$ [which is the case when using both \Eq{eq:C:energy:density} or \Eq{eq:C:H}].
In that case, the condition that two typical trajectories are well-resolved reads
\bea  
\label{eq:collapse:criterion}
R(k) \equiv \frac{\mathbb{E}\left[\left\langle\left(\hat{v}_{\bm k}^s-\left\langle \hat{v}_{\bm k}^s\right\rangle\right)^2\right\rangle\right]}{\mathbb{E}\left(\left\langle \hat{v}_{\bm k}^s\right\rangle ^2\right)}  \ll 1\, .
\eea 
In the limit where the CSL terms are absent, it is clear that the collapse cannot happen. In this case, as discussed above, $\left\langle \hat{v}_{\bm k}^s\right\rangle$ remains zero and 
\begin{align}
    \lim_{\gamma \rightarrow 0}R(k)=\infty \, .
\end{align}
Let us also notice that $R(k)$ can be rewritten as $\mathbb{E}\left[\left\langle \hat{v}_{\bm{k}}^s{}^2 \right\rangle\right]/P_v(k)-1$ and that, upon inverting this formula, one can obtain an expression of the power spectrum in terms of the quantity $R(k)$, namely
\bea 
P_v(k) = \frac{\mathbb{E}\left[\left\langle \hat{v}_{\bm{k}}^s{}^2 \right\rangle\right]}{1+R(k)}\, ,
\eea 
which can be interpreted as follows. In the ``standard'' picture, \ie when the Copenhagen interpretation is used together with the standard Schr\"odinger equation, the power spectrum is simply given by the Born rule
\bea
\label{eq:PowerSpectrun:Copenhagen}
\left. P_v(k)\right\vert_{\mathrm{Copenhagen}} = \left\langle \hat{v}_{\bm{k}}^s{}^2\right\rangle_{\gamma=0}\, .
\eea 
As a consequence, the ``standard'' result is recovered if (i) the state is collapsed, $R(k)\ll 1$, and (ii) $\left\langle \hat{v}_{\bm{k}}^s{}^2 \right\rangle$ is not substantially modified compared to the situation without CSL corrections, $\left\langle \hat{v}_{\bm{k}}^s{}^2 \right\rangle\simeq\left\langle \hat{v}_{\bm{k}}^s{}^2 \right\rangle_{\gamma\simeq 0}$. 
The calculation of \Refa{Gundhi:2021dkh} shows that, when the collapse operator is the Hamiltonian density, the second condition is satisfied, but below, we will show that the first one is not.

\subsection{Which collapse operator?}
\label{subsec:calculationps}

Having established the two conditions needed in order to have a convincing explanation for the emergence of structures in our universe, we now examine which types of collapse operator can satisfy them.  We have seen before that this requires the calculation of two quantities, namely $\mathbb{E}(\left\langle \hat{v}_{\bm{k}}^s\right\rangle^2)$
and $\mathbb{E}\left(\left\langle \hat{v}_{\bm{k}}^s{}^2\right\rangle \right)$.

Let us start with the quantity $\mathbb{E}\left(\left\langle \hat{v}_{\bm{k}}^s{}^2\right\rangle \right)$. From the CSL Schr\"odinger equation, understood in the It\^o formalism, it is shown in  \Refa{Martin:2019jye} that one can derive a Lindblad equation for the statistical average of the density matrix. In turn, this gives rise to a differential equation for the stochastic expectation value of the quantum mean of any operator $\hat{O}$, $\mathbb{E}(\langle \hat{O}\rangle )$. Using this method, it is demonstrated in \Refa{Martin:2019jye} that $\mathbb{E}\left(\left\langle \hat{v}_{\bm{k}}^s{}^2\right\rangle \right)$ obeys a third-order linear differential equation. This equation can then be solved perturbatively and this leads to a solution of the form
 \begin{align}
    \mathbb{E}\left(\left\langle \hat{v}_{\bm{k}}^s{}^2\right\rangle\right)=\left\langle \hat{v}_{\bm{k}}^s{}^2\right\rangle_{\gamma=0}+
    \gamma \, \delta \left\langle \hat{v}_{\bm{k}}^s{}^2\right\rangle_{\mathrm{CSL}}+{\cal O}(\gamma^2).
    \end{align}
In \Refa{Gundhi:2021dkh}, the same strategy is used although the perturbative method appears to be different than the one used in \Refa{Martin:2019jye}.

Notice that this method does not allow us to calculate quantities such as $\mathbb{E}(\langle \hat{O}\rangle^n )$ for $n\neq 1$. This is why the calculation of $\mathbb{E}(\langle \hat{v}_{\bm{k}}^s\rangle^2 )$ must proceed differently. For any quantum operator $\hat{O}$, an equation of motion for its quantum expectation value can be obtained by differentiating $\langle \hat{O} \rangle = \langle \Psi \vert \hat{O} \vert \Psi\rangle$ and using the CSL Schr\"odinger equation~\eqref{eq:csl:Fourier}. One obtains
\begin{align}
\label{eq:eom:meanO}
 \frac{\dd \left\langle \hat{O} \right\rangle}{\dd\eta}  = & \int_{\mathbb{R}^{3+}} \dd^3 \bm{k} \left( - 2i \left\langle \left[\hat{O},\hat{\calH}({\bm{k}})\right]\right\rangle
+\frac{\gamma}{m_0}\xi_{\bm{k}}(\eta) 
\left\langle
\left\lbrace \hat{O},\hat{C}^\dagger_{\bm{k}}-\left\langle \hat{C}^\dagger_{\bm{k}} \right\rangle 
\right\rbrace
\right\rangle
\right. \nonumber \\ & \left.
+ \frac{\gamma}{m_0}\xi^*_{\bm{k}}(\eta)\left \langle\left\lbrace
\hat{O},\hat{C}_{\bm{k}}-\left\langle \hat{C}_{\bm{k}} \right\rangle\right\rbrace\right \rangle
\right. \nonumber  \\  & \left. 
-\frac{\gamma}{2m_0^2} (2\pi)^{3/2}G_{\bm{k}}\left\langle
\left\lbrace \hat{O} , \left[\hat{C}_{\bm{k}}-\left\langle \hat{C}_{\bm{k}}\right\rangle\right]\left[\hat{C}^\dagger_{\bm{k}}-\left\langle \hat{C}^\dagger_{\bm{k}}\right\rangle\right] +
\right. \right. \right. \nonumber \\  & \left. \left. \left.
\left[\hat{C}^\dagger_{\bm{k}}-\left\langle \hat{C}^\dagger_{\bm{k}}\right\rangle\right]\left[\hat{C}_{\bm{k}}-\left\langle \hat{C}_{\bm{k}}\right\rangle\right]\right\rbrace
\right\rangle \right) \, .
\end{align}
In this expression $[,]$ represents the commutator and $\lbrace , \rbrace$ the anticommutator. Let us now consider the case where $\hat{O}$ only involves odd powers of the field space variables in Fourier space, \ie
\begin{align}
\hat{O} = \sum_{n,m}\alpha_{n,m}\hat{v}_{\bm{k}_1}\cdots \hat{v}_{\bm{k}_n} \hat{p}_{\bm{q}_1}\cdots \hat{p}_{\bm{q}_m}\quad \text{with}\quad n+m\quad \text{odd},
\end{align}
(the operators can always be re-ordered in that way upon using canonical commutation relations), where $\alpha_{n,m}$ are unspecified coefficients. In that case, if $\hat{C}_{\bm{k}}=\hat{\calH}_{\bm{k}}$, the right-hand side of \Eq{eq:eom:meanO} only involves terms proportional to expectation values of odd powers of the field variables. Indeed, since the Hamiltonian is quadratic in the field variables, the first term is of order $n+m$, the second term has contributions of order $n+m+2$ and $n+m$ times terms of order $2$, and the last term has contributions of order $n+m+4$, $n+m+2$ times terms of order $2$, and $n+m$ times terms of order $2$. One should then note that since the initial state is set to the Bunch-Davies vacuum, it is Gaussian and separable [in the sense of \Eq{eq:Psi:Fourier}]. As a consequence, all odd moments computed on the initial state vanish. Since odd moments are only sourced by odd moments, we conclude that they vanish at any time, $\langle \hat{O} \rangle = 0$. This property was also noticed in \Refa{Gundhi:2021dkh}, see the remark after Eq.~(92) ``{\it However, for
a collapse operator which is quadratic in the perturbations, and hence in the creation and annihilation operators, one
has that the CSL contribution to $\langle \hat{\cal R}\rangle$ is zero, as one can easily deduce from explicit substitution in Eq.~(87)}''. Concretely, if one takes $\hat{O} = \hat{v}_{\bm{k}}^s$, then one obtains
\bea
\label{eq:<v>:C=H}
\left\langle  \hat{v}_{\bm{k}}^s \right\rangle = 0\, .
\eea 
It follows that the two-point correlation (and, in fact, all statistical moments, see \Sec{sec:NG}) of $\langle  \hat{v}_{\bm{k}}^s \rangle$ vanish. In other words, since all realisations of the stochastic Schr\"odinger equation remain centred around zero, the state never collapses and the power spectrum vanishes,
\begin{align}
P_v(k)=0\, .
\end{align}
Another implication of the previous considerations is that, since the state never collapses, one has 
\begin{align}
R(k)=\infty\, ,
\end{align}
in obvious contradiction with the requirement $R\ll 1$.

This result can be interpreted as a direct consequence of a parity symmetry enjoyed by the free Hamiltonian, which is invariant under the transformation $v(\bm{x}) \to -v(\bm{x})$ [which implies $p(\bm{x}) \to -p(\bm{x})$]. In the standard quantum-mechanical theory, this guarantees that the wavefunctional remains even, $\Psi[v] = \Psi[-v]$. If the collapse operator is the free Hamiltonian, it also enjoys the symmetry, hence the state is still invariant under flipping the sign of $v$, and can thus only be centred on $\langle{v}\rangle=0$. A successful collapse can only be achieved by a collapse operator that breaks some of the symmetries enjoyed by the quantum state in the free theory~\cite{Castagnino:2014cpa}. 

Crucially, it is easy to realise that the above considerations go well beyond the choice $\hat{C}_{\bm k}=\hat{{\cal H}}_{{\bm k}}$ and are in fact true for any collapse operator enjoying the parity symmetry, namely containing only even powers of field variables. As a consequence, we have in fact shown that the state does not collapse (and the power spectrum vanishes) for all collapse operators belonging to this category, a far-fetching conclusion indeed since this allows us to exclude a large class of collapse operators.

Let us now consider the situation where the collapse operator contains odd powers of field variables [as in the case for the energy density~(\ref{eq:C:energy:density})]. Then the second term in the right-hand side of \Eq{eq:eom:meanO}, \ie the one proportional to $\xi_{\bm{k}}(\eta)$, involves expectation values of even powers while the third term involves only odd powers. As a consequence, although, on average, the symmetry is restored, \ie $\mathbb{E}(\langle \hat{v}_{\bm k}^s\rangle)=0$, it is not verified by each stochastic realisation of the wavefunction individually, which guarantees that $\mathbb{E}(\langle \hat{v}_{\bm k}^s\rangle^2)\neq 0$, that is to say a non-vanishing power spectrum. This also implies that the quantity $R(k)$ is no longer divergent. This is why having odd powers in the collapse operator is necessary. However, this is clearly not sufficient: one still needs to check that the collapse operator leads to a power spectrum that is compatible with the data and to a quantity $R(k)$ that is sufficient small. But, at least, we are guaranteed to have $P_v(k)\neq 0$ and $R(k)\neq \infty$.

Among the a priori possible (odd) collapse operators is the energy density~(\ref{eq:C:energy:density}), which was studied in \Refa{Martin:2019jye}. In this article, it was shown that the collapse criterion~\eqref{eq:collapse:criterion} is always satisfied for the values of $\lambda$ and $r_\uc$ compatible with laboratory experiments and quoted in \Sec{sec:intro}. Finding a general expression of $R(k)$ is a non trivial task, but, under the assumption that $R(k)\ll 1$ (implying that the limit $\gamma \rightarrow 0$ cannot be taken in the expressions below), one finds
\bea
\label{eq:Renergy}
R(k)\simeq \begin{cases}
\displaystyle
\frac{1}{1+3456 \dfrac{\gamma\Mp^2 H_\uinf^2}{m_0^2}\ee^{7 \Delta N_*(k)}}+\order{\gamma^2}, \quad & \text{if} \quad H_\uinf r_\uc<\ee^{\Delta N_*(k)},\\
\displaystyle
\dfrac{1}{1+\dfrac{21792}{11} \dfrac{\gamma\Mp^2 H_\uinf^2}{m_0^2}\ee^{14 \Delta N_*(k)} \left(H_\uinf r_\uc\right)^{-7}}+\order{\gamma^2}, \quad & \text{if} \quad H_\uinf r_\uc>\ee^{\Delta N_*(k)}.
\end{cases}
\eea 
We see that the value of $R(k)$ depends on whether the mode under consideration crosses out $r_\uc$ during inflation or during the subsequent radiation era, \ie whether $r_\uc < \ee^{\Delta N_*(k)}/H_\uinf$ or $r_\uc > \ee^{\Delta N_*(k)}/H_\uinf$. In these expressions, which assume that the energy density is not evaluated in the comoving threading, $H_\uinf$ is the value of the Hubble parameter during inflation, and $\Delta N_*(k)$ is the number of \efolds~spent outside the Hubble radius during inflation. It is typically of order $50$ for the scales probed in the CMB. The condition $R(k)\ll 1$ thus imposes 
a lower bound on $\gamma$ (or $\lambda$). Using the central value $r_\uc\sim 10^{-5}$ m, and since inflation must proceed before big-bang nucleosynthesis (so $\sqrt{\Mp H_\uinf} > 10 \MeV$), one finds that collapse always occurs if $\lambda > 10^{-161}\, \mathrm{s}^{-1}$, and since laboratory systems impose $\lambda > 10^{-19}\, \mathrm{s}^{-1}$, it is clear that the collapse is very effective in the early universe.

Unfortunately, unless one makes the specific choice where the energy density is evaluated on the comoving threading, the corrections to the power spectrum are too large. Indeed, as shown in \Refa{Martin:2019jye}, one finds\footnote{When the predicted power spectrum is larger than one, cosmological perturbation theory breaks down and the result cannot be trusted. However, although the precise amplitude of the power spectrum cannot be estimated in that case, it must be large (otherwise a perturbative result would be obtained), which is excluded by observations. Therefore, contrary to what is suggested in the concluding paragraph of \Refa{Gundhi:2021dkh}, \Eq{eq:Pv:rho:result} can be safely used to derive an upper bound on the value of $\gamma$ (or $\lambda$).}
\bea
\label{eq:Pv:rho:result}
P_v(k) = \frac{\left. P_v(k)\right\vert_\mathrm{Copenhagen}}{1+R(k)}
\times
\begin{cases}
1+448 \dfrac{\gamma}{m_0^2}\Mp^2 H_\uinf^2 \epsilon_1 \ee^{\Delta N_*(k)}\ \  & \text{if} \ \  H_\uinf r_\uc<\ee^{\Delta N_*(k)},\\
1+\dfrac{35408}{143}\dfrac{\gamma}{m_0^2}\dfrac{\Mp^2 H_\uinf^2 \epsilon_1 }{(H r_\uc)^{9}} \ee^{10 \Delta N_*(k)}\ \ & \text{if} \ \  H_\uinf r_\uc>\ee^{\Delta N_*(k)},
\end{cases}
\eea 
where $\left. P_v(k)\right\vert_\mathrm{Copenhagen}$ denotes the ``standard'' result~\eqref{eq:PowerSpectrun:Copenhagen}, which we stress again is in excellent agreement with observations, and $\epsilon_1$ is the first slow-roll parameter. At the central value $r_\uc\sim 10^{-5}$ m, for the corrections to the standard result to remain negligible, one needs to impose $\lambda<10^{-73}\ \mathrm{s}^{-1}$ if inflation proceeds at $H_\uinf=10^{-5}\Mp$, which excludes the values in agreement with other laboratory experiments. As already mentioned, only if the energy density is evaluated on the comoving threading can the theory be made compatible with CMB measurements.

Let us note that those considerations assume that the values of $\lambda$ and $r_\uc$ are the same during inflation and in laboratory experiments. However, in a relativistic context, one may expect these parameters to run with the energy at which the experiment is being performed, and to assume different values at the high energies at which inflation proceeds from those constrained in laboratory setups. Such a running would have to account for at least 54 order of magnitude in $\lambda$, which may seem unlikely, but without a fully relativistic formulation of CSL, this remains a possibility.

\section{Constraining the collapse operator with non-Gaussianities}
\label{sec:NG}

So far, we have shown that even collapse operators are ruled out and that only collapse operators that contain odd powers of field variables can successfully collapse the wavefunction of cosmological fluctuations. In this category, the sub-class of collapse operators that are linear in field variables was studied more extensively in \Refa{Martin:2019jye} and, in that article, it was demonstrated that if the collapse operator is taken as the energy density, unless it is evaluated in the comoving threading, it leads to predictions that are incompatible with the current CMB data, for the values of the CSL parameters that are allowed by laboratory experiments.

However, we have not addressed the possibility that the collapse operator is dominated by other odd powers of field variables, say cubic powers. In fact, even in the case where the collapse operator is taken as the Hamiltonian density, it contains cubic and higher-order terms~\cite{Maldacena:2002vr}, that might be able to lead to the collapse of the wavefunction.\footnote{Since the amplitude of higher-order terms is Planck-suppressed, and given that the leading, quadratic terms are unable to substantially alter the width of the wavefunction (see  \Refa{Gundhi:2021dkh}), it may seem unlikely that higher-order terms make the wavefunction collapse, although it would remain to be checked explicitly.} In that case, as we shall now see, non-Gaussianities can further tighten the choice of the collapse operator. Current measurements of the CMB impose the statistics of cosmological fluctuations to be Gaussian up to tightly constrained deviations~\cite{Akrami:2019izv}. Therefore, it is not sufficient that the two-point correlation function of $\left \langle \hat{v}_{\bm k}^s\right \rangle $ is compatible with measurements of the power spectrum, one must also ensure that the whole statistics of $\left \langle \hat{v}_{\bm k}^s\right \rangle $ is Gaussian or quasi Gaussian. 

Another reason why studying the statistics of $\left \langle \hat{v}_{\bm k}^s\right \rangle $ is important is that, when the collapse operator is taken as the energy density in the comoving threading, one may wonder whether or not the non-Gaussianity test can be passed (since the power-spectrum test is, see above).
 Interestingly enough, we will find that if the collapse operator is linear in field variables, the full statistics of $\left \langle \hat{v}_{\bm k}^s\right \rangle $ can be determined exactly. 

For explicitness, we consider the case where the (quadratic) Hamiltonian is the one of a parametric oscillator, 
\bea
\label{eq:Hamiltonian:parametric:oscillator}
\hat{H}=\int_{\mathbb{R}^{3+}}\dd\bm{k} \left[\hat{p}_{\bm{k}}+\omega^2\left(k,\eta\right)\hat{v}_{\bm{k}}^2\right] ,
\eea 
since it is the case for cosmological perturbations at leading order in cosmological perturbation theory, but the techniques we present here can be easily generalised to any quadratic Hamiltonian [which can otherwise always be cast in the form~\eqref{eq:Hamiltonian:parametric:oscillator} upon performing a suitable canonical transformation~\cite{Grain:2019vnq}]. 

As shown in \Sec{sec:cslinflation}, with the choice~\eqref{eq:C:linear}, the wavefunction remains factorisable in Fourier space, see \Eq{eq:Psi:Fourier}, and the CSL equation admits Gaussian solutions of the form
\bea
\label{eq:SingleGaussianR}
\Psi_{\bm{k}}^s\left(\eta,v_{\bm{k}}^s\right)=
\vert N_{\bm{k}}\left(\eta\right)\vert \exp\Bigl\lbrace &
-\Rea  \Omega_{\bm{k}}\left(\eta\right)
\left[v_{\bm{k}}^s-\bar{v}_{\bm{k}}^s\left(\eta\right)\right]^2
+i\sigma_{\bm k}^s(\eta)+i\chi_{\bm k}^s(\eta)
v_{\bm{k}}^s
 \\ &
-i\Ima  \Omega_{\bm k}(\eta )
\left(v_{\bm{k}}^s\right)^2\Bigr\rbrace\, ,
\eea
where, for the state to be normalised, one has
\begin{align}
\label{eq:Norm:WaveFunction}
\vert N_{\bm{k}}\vert =\left(\frac{2\Rea  \Omega_{\bm{k}}}{\pi}\right)^{1/4}\, .
\end{align}
This can be seen by plugging \Eq{eq:SingleGaussianR} into the CSL equation~\eqref{eq:csl:Fourier}, and by checking that \Eq{eq:SingleGaussianR} indeed gives a solution provided the parameters of the Gaussian obey the following equations of motion~\cite{Martin:2019jye},
\begin{align}
\label{eq:eom:N}
\frac{\dd \ln\left\vert N_{\bm{k}}\left(\eta\right)\right\vert}{\dd\eta} 
 = & \frac{1}{4  \Rea  \Omega_{\bm{k}}}
\frac{\dd \Rea  \Omega_{\bm{k}}}{\dd\eta},\\
%%%
\frac{\dd \Rea  \Omega_{\bm{k}}}{\dd\eta} =& 
\frac{\gamma}{m_0^2} a^4\alpha_{\bm k}^2-4\frac{\gamma}{m_0^2} a^4\beta_{\bm k}^2 
\left[\left(\Rea  \Omega_{\bm{k}}\right)^2
-\left(\Ima  \Omega_{\bm{k}}\right)^2\right]
+4 \Rea  \Omega_{\bm{k}} \Ima  \Omega_{\bm{k}}
\nonumber \\ &
-4\frac{\gamma}{m_0^2} a^4 \alpha_{\bm k} \beta_{\bm k} \Ima  \Omega_{\bm{k}},\\
\label{eq:eom:OmegaR}
%%%
\frac{\dd \Ima  \Omega_{\bm{k}}}{\dd\eta} =& 
\frac{1}{2}\omega^2(k,\eta)-2\left[\left(\Rea  \Omega_{\bm{k}}\right)^2
-\left(\Ima  \Omega_{\bm{k}}\right)^2\right]
-8\frac{\gamma}{m_0^2} a^4 \beta_{\bm k}^2  \Rea  \Omega_{\bm{k}} 
\Ima  \Omega_{\bm{k}}
\nonumber \\ &
+4\frac{\gamma}{m_0^2} a^4 \alpha_{\bm k} \beta_{\bm k}  
\Rea  \Omega_{\bm{k}},\\
\label{eq:eom:OmegaI}
%%%
\frac{\dd \bar{v}_{\bm{k}}^s}{\dd\eta} = &\chi_{\bm{k}}^s
-2\bar{v}_{\bm{k}}^s\Ima  \Omega_{\bm{k}}
+\frac{\sqrt{\gamma} a^2}{2m_0\Rea  \Omega_{\bm{k}}}
\left(\alpha_{\bm k} - 2 \beta_{\bm k} \Ima  \Omega_{\bm{k}}\right)
\bar{\xi}_{\bm{k}}^s(\eta),\\
\label{eq:eom:vbar}
%%%
\frac{\dd\chi_{\bm{k}}^s}{\dd\eta}= & 2 \Ima  \Omega_{\bm{k}}\chi_{\bm{k}}^s
- 4 \left(\Rea  \Omega_{\bm{k}}\right)^2\bar{v}_{\bm{k}}^s
+8\frac{\gamma}{m_0^2} a^4\beta_{\bm k} \Rea  \Omega_{\bm{k}} \bar{v}_{\bm{k}}^s 
\left(\alpha_{\bm k}-2\beta_{\bm k} \Ima  \Omega_{\bm{k}}\right)
\nonumber \\ &
+2\frac{\sqrt{\gamma}}{m_0}a^2\beta_{\bm k} \Rea  \Omega_{\bm{k}}
\bar{\xi}_{\bm{k}}^s(\eta),\\
\label{eq:eom:chi}
%%%%
\frac{\dd\sigma_{\bm{k}}^s}{\dd\eta} = & -\Rea  \Omega_{\bm{k}}
+2\left(\Rea  \Omega_{\bm{k}}\right)^2(\bar{v}^s_{\bm{k}})^2
-\frac{(\chi^s_{\bm{k}})^2}{2}
-2\frac{\sqrt{\gamma}}{m_0}a^2\beta_{\bm k} \Rea  \Omega_{\bm{k}} \bar{v}_{\bm k}^s
\bar{\xi}_{\bm{k}}^s(\eta)
\nonumber \\ &
+\frac{\gamma a^4}{2m_0^2}\beta_{\bm k}
\left(\alpha_{\bm k}-2\beta_{\bm k}\Ima  \Omega_{\bm{k}}\right)
\left(1-8\Rea  \Omega_{\bm{k}} \bar{v}_{\bm{k}}^2\right)
. 
\end{align}
This system comprises six coupled, non-linear and stochastic differential equations, and is therefore a priori difficult to study. However, this apparent complicated structure does not prevent one from solving the system as follows. 

The first equation, \Eq{eq:eom:N}, is solved by \Eq{eq:Norm:WaveFunction}, and simply guarantees that the norm of the wavefunction is preserved.

The second and third equations, \Eqs{eq:eom:OmegaR} and~\eqref{eq:eom:OmegaI}, are not stochastic [the noise $\bar{\xi}_\eta(\bm{k})$ does not appear in these equations] and indicate that $\Omega_{\bm{k}}$ decouples from the other parameters of the wavefunction. By combining \Eqs{eq:eom:OmegaR} and~\eqref{eq:eom:OmegaI}, one can indeed derive an autonomous equation for
$\Omega_{\bm{k}} = \Rea \Omega_{\bm{k}} + i \Ima \Omega_{\bm{k}}$,
namely
\bea
\label{eq:eom:Riccatti:Omegak}
 \Omega_{\bm{k}}' = -2\left(i+2\frac{\gamma}{m_0^2} a^4 \beta_{\bm k}^2\right)
\Omega_{\bm{k}}^2+4i\frac{\gamma}{m_0^2} a^4 
\alpha_{\bm k}\beta_{\bm k} \Omega_{\bm{k}} 
+\frac{\gamma}{m_0^2} a^4\alpha_{\bm k}^2 + \frac{i}{2}\omega^2(k,\eta)\, .
\eea
This is a first-order, non-linear differential equation known as a Riccati equation, and it can be cast in terms of a second-order, linear differential equation by introducing the change of variables~\cite{Martin:2019jye}
\begin{align}
\label{eq:f:g:redef}
\Omega_{\bm{k}} = \frac{1}{2\left(i+2\gamma a^4 \beta_{\bm k}^2/m_0^2\right)} 
\left(\frac{g_{\bm k}'}{g_{\bm k}}-\frac12 C_1\right)\, ,
\end{align}
where the function $g_{\bm k}(\eta)$ obeys
\begin{align}
\label{eq:exactg}
g_{\bm k}''+\left(-\frac12 C_1'-\frac14C_1^2+C_2\right)g_{\bm k}=0,
\end{align}
and the coefficients $C_1$ and $C_2$ are given by 
\bea
\label{eq:defC12}
C_1&\equiv - 2 i \frac{\gamma}{m_0^2} \left[2 a^4 \alpha_{\bm k} \beta_{\bm k}
-\frac{\left(a^4 \beta_{\bm k}^2\right)'}{1-2 i\gamma a^4 \beta_{\bm k}^2/m_0^2}
\right], \\
C_2 & \equiv \left(1-2i\frac{\gamma}{m_0^2} a^4 \beta_{\bm k}^2\right)
\left[\omega^2(k,\eta)-2i\frac{\gamma}{m_0^2} a^4 \alpha_{\bm k}^2\right] .
\eea
Let us note that, when $\gamma=0$, $-\frac12 C_1'-\frac14C_1^2+C_2=\omega^2$ and one recovers the standard, classical equation of motion for $v_{\bm{k}}^s$.
In this way, one obtains the parameter $\Omega_{\bm{k}}(\eta)$, hence the parameter $N_{\bm{k}}(\eta)$ using \Eq{eq:Norm:WaveFunction}. From the form of the wavefunction~\eqref{eq:SingleGaussianR}, one can show that
\bea
 \left\langle \left(\hat{v}_{\bm{k}}^s-\left\langle \hat{v}_{\bm{k}}^s \right\rangle\right)^2
\right\rangle = \frac{1}{4 \Rea \Omega_{\bm{k}}}\, .
 \eea
This means that this quantum expectation value, which is nothing but the width of the wave-function, is a non-stochastic quantity (which justifies the statement made in footnote~\ref{footnote:Deterministic:Variance}), and can be obtained from the above considerations. 

Having determined $\Omega_{\bm{k}}(\eta)$, one notices that the two next equations in the system, \Eqs{eq:eom:vbar} and~\eqref{eq:eom:chi}, form a linear autonomous subsystem for $\bar{v}_{\bm{k}}^s$ and $\chi_{\bm{k}}^s$. Upon introducing the vector $\bm{X}=(\bar{v}_{\bm{k}}^s,\chi_{\bm{k}}^s)^\mathrm{T}$, where $\mathrm{T}$ denotes the transpose, they can be written in matricial form as
\bea 
\label{eq:Langevin:X}
\frac{\dd \bm{X}}{\dd\eta} = \bm{A} \cdot \bm{X} + \bm{Y} \bar{\xi}_{\bm{k}}^s(\eta)\, ,
\eea
where
\bea
\label{eq:bmA:def}
\bm{A}= \left(
\begin{array}{ccc}
\displaystyle
-2\Ima\Omega_{\bm{k}} &\quad & 1\\ \\
\displaystyle
-4\left(\Rea\Omega_{\bm{k}}\right)^2 + 8 \frac{\gamma}{m_0^2} a^4 \beta_{\bm{k}} \Rea\Omega_{\bm{k}} \left(\alpha_{\bm{k}}-2\beta_{\bm{k}}\Ima\Omega_{\bm{k}}\right) &\quad &
2 \Ima\Omega_{\bm{k}}
\end{array}
\right)
\eea 
and
\bea 
\bm{Y}
= \left(
\begin{array}{c}
\displaystyle
\frac{\sqrt{\gamma} a^2}{2m_0\Rea  \Omega_{\bm{k}}}
\left(\alpha_{\bm k} - 2 \beta_{\bm k} \Ima  \Omega_{\bm{k}}\right)\\ \\
\displaystyle
2\frac{\sqrt{\gamma}}{m_0}a^2\beta_{\bm k} \Rea  \Omega_{\bm{k}}
\end{array}
\right)\, .
\eea
The Langevin equation~\eqref{eq:Langevin:X} gives rise to a Fokker-Planck equation~\cite{Risken:1984book} for the probability density associated the vector $\bm{X}$,
\bea 
\label{eq:fokker}
\frac{\partial P\left(\bm{X},\eta\right) }{\partial\eta} = -\displaystyle\sum_{i,j=1}^2 A_{ij} \frac{\partial}{\partial X }_i\left[ X_j P(\boldsymbol{X},\eta)\right]+\frac{1}{2}\displaystyle\sum_{i,j=1}^2
Y_i Y_j
\frac{\partial^2 P(\boldsymbol{X},\eta)}{\partial X_i\partial X_j}\, .
\eea 
Since the dynamics of $\bm{X}$ is linear, it can be solved by making use of the Green's matrix formalism, as shown in detail in \Refa{Grain:2017dqa} (in a different context). The Green's matrix $\boldsymbol{G}(\eta,\eta_0)$ is defined as the $2$ by $2$ matrix that is a solution of the homogeneous (hence deterministic) problem associated to the stochastic dynamics of $\bm{X}$, \ie 
\bea
\label{eq:eom:Green}
\frac{\partial \boldsymbol{G}(\eta,\eta_0)}{\partial\eta}=\boldsymbol{A}(\eta)\boldsymbol{G}(\eta,\eta_0)+\boldsymbol{I}\delta(\eta-\eta_0),
\eea
where $\boldsymbol{I}$ is the $2$ by $2$ identity matrix. 
It can be constructed explicitly from solutions of the linear homogeneous system $\dd{\boldsymbol{X}}/\dd\eta=\boldsymbol{A}(\eta)\boldsymbol{X}$. Let us indeed assume that two independent solutions $(\bar{v}_{\bm{k}}^{s\,(1)},{\chi_{\bm{k}}^{s\,(1)}})$ and $({\bar{v}_{\bm{k}}^{s\,(2)}},{\chi_{\bm{k}}^{s\,(2)}})$ of this linear homogeneous system are known. The so-called ``fundamental'' matrix of the system is defined as
\bea
\label{eq:U:def}
\boldsymbol{U}(\eta)=\left(\begin{array}{cc}
	{\bar{v}_{\bm{k}}^{s\,(1)}}(\eta) & {\chi_{\bm{k}}^{s\,(1)}}(\eta) \\
	{\bar{v}_{\bm{k}}^{s\,(2)}}(\eta) & {\chi_{\bm{k}}^{s\,(2)}}(\eta)
\end{array}\right).
\eea
By construction, one can check that $\dd \boldsymbol{U}(\eta)/\dd \eta =\boldsymbol{A}(\eta)\boldsymbol{U}(\eta)$. Let us also notice that since the two solutions are independent, $\det(\boldsymbol{U})\neq0$. The matrix $\boldsymbol{U}$ is then invertible and gives rise to the Green's matrix
\bea
\label{eq:Green:fundamental}
	\boldsymbol{G}(\eta,\eta_0)&=&\boldsymbol{U}(\eta)\left[\boldsymbol{U}(\eta_0)\right]^{-1}
	\Theta(\eta-\eta_0) ,
\eea
which satisfies \Eq{eq:eom:Green}.\footnote{One can also note that ${\dd}\det[\boldmathsymbol{U}(\eta)]/{\dd\eta}=\mathrm{Tr}\left[\boldmathsymbol{A}(\eta)\right]\det[\boldmathsymbol{U}(\eta)]$ with ``$\mathrm{Tr}$'' being the trace operation. The coefficients matrix $\boldsymbol{A}$, defined in \Eq{eq:bmA:def}, is traceless and $\det[\boldsymbol{U}(\eta)]$ is thus a conserved quantity. It is therefore sufficient to find two solutions such as $\det[\boldsymbol{U}(\eta_0)]\neq0$, and this ensures the Green's matrix to be properly defined throughout the evolution.}

It is worth pointing out that the search for the two independent solutions is a simple task because of the following remark. The first-order equation $\dd{\boldsymbol{X}}/\dd\eta=\boldsymbol{A}(\eta)\boldsymbol{X}$ can be cast in terms of a single, second-order equation for ${\bar{v}_{\bm{k}}^{s\,(i)}}$, namely $({\bar{v}_{\bm{k}}^{s\,(i)}})^{\prime\prime} + \mu^2 {\bar{v}_{\bm{k}}^{s\,(i)}}=0$, where $\mu^2=-(A_{11}^\prime+A_{21}+A_{11}^2)$, where we have used that $A_{11}=-A_{22}$ and where a prime denotes derivation with respect to time $\eta$. Making use of \Eq{eq:eom:OmegaI} to evaluate $A_{11}^\prime$, all terms involving $\gamma$ cancel out, and one obtains $\mu^2=\omega^2$, hence
\bea
\left({\bar{v}_{\bm{k}}^{s\,(1)}}\right)^{\prime\prime} + \omega^2(k,\eta) {\bar{v}_{\bm{k}}^{s\,(1)}}=0\, .
\label{eq:eom:vi:class}
\eea 
This is nothing but the standard, classical equation of motion for $v_{\bm{k}}^s$, and is a particular case of \Eq{eq:exactg} when $\gamma=0$. Analytical solutions to \Eq{eq:eom:vi:class} are known in most cosmological backgrounds (for instance when the equation-of-state parameter is constant, or when inflation proceeds in the slow-roll regime).
Once ${\bar{v}_{\bm{k}}^{s\,(i)}}$ is obtained, one can readily derive $ {\chi_{\bm{k}}^{s\,(i)}}$ from the relation $ {\chi_{\bm{k}}^{s\,(i)}} = ({\bar{v}_{\bm{k}}^{s\,(i)}})^{\prime} +2 \Ima\Omega_{\bm{k}} {\bar{v}_{\bm{k}}^{s\,(i)}}$.

Having determined the Green function, solutions to the Fokker-Planck equation~\eqref{eq:fokker} can be written formally by means of the kernel function $\mathcal{W}(\boldsymbol{X},\eta|\boldsymbol{X}_0,\eta_0)$,
\bea
P\left(\boldsymbol{X},\eta\right)=\displaystyle\int \dd\boldsymbol{X}_0\mathcal{W}(\boldsymbol{X},\eta|\boldsymbol{X}_0,\eta_0)P\left(\boldsymbol{X}_0,\eta_0\right) ,
\eea
where the kernel function has the Gaussian form
\bea
\label{eq:Gaussian:Green}
	\mathcal{W}\left(\boldsymbol{X},\eta|\boldsymbol{X}_0,\eta_0\right)=\displaystyle\frac{1}{\sqrt{2\pi^2\det\left[\boldsymbol{\Sigma}(\tau)\right]}}\exp\left\lbrace-\frac{1}{2}\left[\boldsymbol{X}-\boldsymbol{X}_{\mathrm{det}}(\eta)\right]^\dag\boldsymbol{\Sigma}^{-1}(\eta)\left[\boldsymbol{X}-\boldsymbol{X}_{\mathrm{det}}(\eta)\right]\right\rbrace. 
\eea
In this expression, $\dag$ denotes the conjugate-transpose, and $\boldsymbol{X}_{\mathrm{det}}(\eta)=\boldmathsymbol{G}(\eta,\eta_0)\boldmathsymbol{X}_0$ is the deterministic trajectory that would be obtained in the absence of the noise term and starting from $\boldmathsymbol{X}(\eta_0)=\boldmathsymbol{X}_0$. From \Eq{eq:Gaussian:Green}, assuming initially a distribution given by a Dirac function, one can check that $\mathbb{E}\left[\boldsymbol{X}(\eta)\right]=\int\dd\boldsymbol{X}~\boldsymbol{X}\mathcal{W}(\boldsymbol{X},\eta|\boldsymbol{X}_0,\eta_0)=\boldsymbol{X}_{\mathrm{det}}$, which means that the deterministic trajectory is also the averaged trajectory. Finally, $\boldsymbol{\Sigma}$ is the covariance matrix, which is obtained as the forward propagation of the diffusion matrix,
\bea
\label{eq:Sigma}
	\boldsymbol{\Sigma}(\eta)=\displaystyle\int^\eta_{\eta_0}\dd s~\boldsymbol{G}(\eta,s)\boldsymbol{\mathbb{Y}}(s)\boldsymbol{G}^\dag(\eta,s) ,
	\quad\text{where}\quad \mathbb{Y}_{i j} = Y_i Y_j\, .
\eea	
It is related to the two-point (statistical, not quantum) correlation function of $\bm{X}$ through the following expression
\bea
\mathbb{E}\left(\left\lbrace\boldsymbol{X}(\eta)-\mathbb{E}\left[\boldsymbol{X}(\eta)\right]\right\rbrace\left\lbrace\boldsymbol{X}(\eta)-\mathbb{E}\left[\boldsymbol{X}(\eta)\right]\right\rbrace^\dag\right)=\boldsymbol{\Sigma}(\eta) .
\eea 
Therefore, we reach the conclusion that $\bm{X}$ follows a Gaussian statistics. In particular, $\bar{v}_{\bm{k}}^s$, which, as argued in \Sec{subsec:define}, corresponds to the quantity of observational interest, has Gaussian statistics. 

Therefore, the above result answers the question asked at the beginning of this section. If the collapse operator is taken to be the energy density (in particular, the energy density in the comoving threading since we saw that this is the only remaining possibility), then the corresponding observable predictions are Gaussian, in agreement with CMB measurements~\cite{Akrami:2019izv}. In that case, non-Gaussianities only arise through higher-order terms in cosmological perturbation theory, as in the standard calculation. This is an important result that confirms that the energy density in the comoving threading is a consistent candidate.

What about the other (non-linear) odd collapse operators? Given the previous considerations, it is clear that the Gaussian statistics of  $\bar{v}_{\bm{k}}^s$ arises because the collapse operator is linear. If the collapse operator is odd but not linear, the statistics of $\bar{v}_{\bm{k}}^s$ will not be Gaussian, and mode coupling will become important too, for a similar argument as the one given around \Eq{eq:H_k}. 
Then, given the non-perturbative nature of the collapse process, it seems likely that this extra source of non-Gaussianity will be in conflict with the current astrophysical data. 
This constitutes a fundamental difference between linear and non-linear collapse operators, which allows us to discard the later possibility. 

\section{Conclusion}
\label{sec:Discussion}
In this work we have discussed the conditions a collapse operator must satisfy in order to properly describe the emergence of cosmological structures in the early universe. We have found that if the collapse operator is even in the field variables, it is invariant under their sign flipping, hence the collapse theory is endowed with the same symmetry and all the realisations of the wavefunctional remain centred around a vanishing configuration. 
In this case, collapse does not occur, neither in the early universe nor later, and the theory is ruled out.

This is the case of the proposal made in \Refa{Gundhi:2021dkh} where the collapse operator is taken as the free Hamiltonian, which is quadratic in field variables, hence unsuitable for cosmology. One may argue that, at higher order in cosmological perturbation theory, the free Hamiltonian also contains Planck-suppressed cubic (and higher-order) powers of the fields. However, even if the amplitude of those suppressed terms were large enough to lead to the collapse of the wavefunction, we have argued that they would lead to non-Gaussian outcomes, with substantial mode coupling, likely in contradiction with observations. 

In contrast, we have found that if the collapse operator is linear in the field variables, not only does it have the potential to make the wavefunction collapse (since it is odd), but it also leads to outcomes that are distributed according to a Gaussian law. The main conclusion of this work is therefore that, in cosmology, the collapse operator must be linear (at leading order) in the field variables.

Having said this, the next question is of course which linear collapse operator should we take, and how is it connected to the non-relativistic limit of the theory, where the collapse operator is the mass-density operator. In \Refa{Martin:2019jye}, we have proposed the energy density as a natural extension of the notion of mass density. It is linear in the field variables at leading order so it is a priori a good candidate, according to the considerations presented in this work. However, unless it is evaluated in the comoving threading, we have shown in \Refa{Martin:2019jye} that, although it is very efficient at collapsing the wavefunction, it does not produce a quasi scale-invariant power spectrum as observed in CMB measurements. In \Refa{Martin:2019jye}, we have generalised those results to the case where the collapse operator is built from contractions of the stress-energy tensor, as proposed in \Refs{Bengochea:2020efe, Bengochea:2020qsd}.

At this stage, two possibilities therefore remain. Either the collapse operator is the energy density in the comoving threading, either it is another linear combination of the field variables, that has the same dimension as the energy density but that cannot be interpreted as such. This important conclusion may serve as a useful guide in the attempt to extend collapse theories to the relativistic frameworks, since it imposes properties of the collapse operator in the cosmological setup.

\acknowledgments We would like to thank Anirudh Gundhi,
Jos\'e Luis Gaona-Reyes, Matteo Carlesso and Angelo Bassi for
interesting comments and discussions.

\bibliographystyle{JHEP}
\bibliography{CSLwithH0.bib}
\end{document}